\documentclass[12pt]{article}
\usepackage{epsf} 

\begin{document}

\newcommand{\sheptitle}
{ Scale Invariance of Dirac Condition $g_e g_m = 1$ in Type
$0$  String Approach to Gauge Theory }

\newcommand{\shepauthor}
{
Ian I.~Kogan$^{a,}$\footnote[1]{i.kogan@physics.ox.ac.uk}, 
Gloria Luz\'on$^{a,b,}$\footnote[2]{luzon\,@\,thphys.ox.ac.uk ,
 luzon\,@\,posta.unizar.es }}

\newcommand{\shepaddress}
{
$a$ Dept. of Physics, Theoretical Physics, University of Oxford,\\
  1 Keble Road, Oxford, OX1 3NP, United Kingdom\\
$b$  Dept. de F\'{\i}sica Te\'orica, Facultad de Ciencias,\\
  Universidad de Zaragoza, 50009 Zaragoza, Spain   }
 
\newcommand{\shepabstract}
{In this letter  we shall discuss  a  description of 
 non-supersymmetric  four-dimensional Yang-Mills theory   based on 
Type $0$ strings recently proposed by Klebanov and Tseytlin. 
The three brane  near-horizon geometry  allows one to
study the UV behaviour of the gauge theory. Following Minahan and
Klebanov and Tseytlin  we  shall discuss how  the
gravity solution  reproduces  logarithmic renormalization of coupling
constant $g_e$ extracted  from  quark-antiquark potential  and then show
that effective coupling constant $g_m$ describing monopole-antimonopole
interactions is of zero-charge  type and Dirac condition $g_e g_m = 1$
is scale invariant in logarithmic approximation.
}

\begin{titlepage}
\begin{flushright}
OUTP-99-10-P\\
hep-th/9902086\\
February 1999

\end{flushright}
\vspace{0.5in}
\begin{center}
{\large{\bf \sheptitle}}
\bigskip \\ \shepauthor \\ \mbox{} \\ {\it \shepaddress} \\ 
\vspace{0.5in}
{\bf Abstract} \bigskip \end{center} \setcounter{page}{0}
\shepabstract
\vspace{0.5in}
\begin{flushleft}
OUTP-99-10-P\\
\end{flushleft}
\end{titlepage}

\def\sspace{\baselineskip = .16in}
\def\dspace{\baselineskip = .30in}
\def\beq{\begin{equation}}
\def\eeq{\end{equation}}
\def\bea{\begin{eqnarray}}
\def\eea{\end{eqnarray}}
\def\bq{\begin{quote}}
\def\eq{\end{quote}}
\def\ra{\rightarrow}
\def\lra{\leftrightarrow}
\def\ups{\upsilon}
\def\bq{\begin{quote}}
\def\eq{\end{quote}}
\def\ra{\rightarrow}
\def\un{\underline}
\def\ov{\overline}
\def\ord{{\cal O}} 

\newcommand{\plb}[3]{{{\it Phys.~Lett.}~{\bf B#1} (#3) #2}}
\newcommand{\npb}[3]{{{\it Nucl.~Phys.}~{\bf B#1} (#3) #2}}
\newcommand{\prd}[3]{{{\it Phys.~Rev.}~{\bf D#1} (#3) #2}}
\newcommand{\ptp}[3]{{{\it Prog.~Theor.~Phys.}~{\bf #1} (#3) #2}}
\newcommand{\ijmpa}[3]{{{\it Int.~J.~Mod.~Phys.}~{\bf A#1} (#3) #2}}
\newcommand{\prl}[3]{{{\it Phys.~Rev.~Lett.}~{\bf #1} (#3) #2}}
\newcommand{\hepph}[1]{{\tt hep-ph/#1}}
\newcommand{\hepth}[1]{{\tt hep-th/#1}}
\newcommand{\grqc}[1]{{\tt gr-qc/#1}} 
\newcommand{\leqsim}{\,\raisebox{-0.6ex}{$\buildrel < \over \sim$}\,}
\newcommand{\geqsim}{\,\raisebox{-0.6ex}{$\buildrel > \over \sim$}\,}
\newcommand{\be}{\begin{equation}}
\newcommand{\ee}{\end{equation}}
\newcommand{\ba}{\begin{eqnarray}}
\newcommand{\ea}{\end{eqnarray}}
\newcommand{\nn}{\nonumber}
\newcommand{\cf}{\mbox{{\em c.f.~}}}
\newcommand{\ie}{\mbox{{\em i.e.~}}}
\newcommand{\eg}{\mbox{{\em e.g.~}}}
\newcommand{\mpl}{\mbox{$M_{pl}$}}
\newcommand{\ol}[1]{\overline{#1}}
\def\gev{\,{\rm GeV }}
\def\tev{\,{\rm TeV }}
\def\dd{\mbox{d}}
\def\etal{\mbox{\it et al }}
\def\half{\frac{1}{2}}
\def\Tr{\mbox{Tr}}
\def\bra{\langle}
\def\ket{\rangle}
\def\lim{\mbox{{\bf L}} }
\def\nlim{\mbox{{\bf NL}} }
\def\sclim{\mbox{\tiny{\bf L}} }
\def\scnlim{\mbox{\tiny{\bf NL}} }
\def\nlimc{\mbox{{\bf NL$_{closed}$}} }
\def\nlimo{\mbox{{\bf NL$_{open}$}} }
\def\Vp{V_{\parallel}}
\def\Vt{V_{\perp}} 
\def\Rt{R_{\perp}}
\newcommand{\smallfrac}[2]{\frac{\mbox{\small #1}}{\mbox{\small #2}}}

\def\CAG{{\cal A/\cal G}}           \def\CO{{\cal O}} \def\CZ{{\cal Z}}
\def\CA{{\cal A}} \def\CC{{\cal C}} \def\CF{{\cal F}} \def\CG{{\cal G}}
\def\CL{{\cal L}} \def\CH{{\cal H}} \def\CI{{\cal I}} \def\CU{{\cal U}}
\def\CB{{\cal B}} \def\CR{{\cal R}} \def\CD{{\cal D}} \def\CT{{\cal T}}
\def\CM{{\cal M}} \def\CP{{\cal P}}
\def\CN{{\cal N}} \def\CS{{\cal S}}

\newpage 
\clearpage

\setcounter{section}{0}
\setcounter{equation}{0}
\def\theequation{\arabic{section}.\arabic{equation}}

\section{Introduction}

Non-perturbative description of Yang-Mills fields  in four dimensions
is still one of the most challenging problems in modern theoretical
physics. One of the most popular views is that in the large $N$ limit
 this description must be based on some kind of string theory. 
 In spite of the huge amount of work on the subject in the
last two decades 
 the proper description of $SU(N)$ gauge theories  in the large
$N$  limit is still an open problem.   Nevertheless, it was a
remarkable progress in the last two years which gave us 
a deeper understanding of non-perturbative aspects of gauge theory  using
  new ideas in string theory.
  The duality of $N=4$ supersymmetric Yang-Mills 
to ten dimensional supergravity on 
$AdS_5\times S^5$ \cite{M1}\cite{P1}\cite{W1} 
has been used to obtain exact results in the large $N$ limit of the
strongly coupled superconformal gauge theory. This theory  has 
 vanishing beta function and  is realized
 as the world-volume theory of $N$ coincident D3-branes of type $IIB$ string 
theory. This set of branes causes the near horizon geometry of 
$AdS_5\times S^5$
and the classical type $IIB$ theory can be approximated 
by the compactified supergravity in the limit $\lambda_{IIB}\rightarrow 0$,  
 $g_{eff}$  large but fixed.

The interaction between charged particles has been analyzed in this
context. 
The massless sources might be delicate to deal with, since their long-range
fields are exponentially suppressed due to conformal invariance. Hence,
the efforts have been concentrated on the study of massive electric and
magnetic particles. The computation of the Nambu-Goto action in an $AdS$
background for a static string configuration allowed Rey and Yee \cite{RY} 
and Maldacena \cite{M2} to find the coulomb potential between quark and
antiquarks at zero temperature. Some results have been also obtained for
finite temperature \cite{RY2}, \cite{W2}, \cite{BISY} where a new
confining branch appears. In \cite{Mi1} Minahan extended the results of
finite temperature to the cases of monopole-antimonopole and monopole-quark 
interactions. Here, he finds an attractive coulomb force for the
monopole-antimonopole pair as a function of the coupling manifestly
dual under the transformation $g_e\rightarrow 1/ g_m$. Therefore, the Dirac
condition corresponding to the gauge theory has arised as a result of
stringy computations.

The number of interesting predictions due to the AdS/CFT conjecture has
led to do research into a possible extension to the non-supersymmetric
case. By heating up a maximally supersymmetric gauge theory all the
supersymmetries get broken. In this approach the gauge theory is dual to
near-extremal branes whose near-horizon geometry corresponds to a black
hole in AdS space \cite{W2}\cite{BH}. Recently, a different approach was 
suggested by Polyakov  \cite{P2} : the Type $0$ string theory
\cite{type0} in $d\le 10$  could be
used to extend the AdS/CFT duality to non-supersymmetric non-conformal field
 theories \cite{P2}. This idea inspired the conjecture of Klebanov and
Tseytlin: the existence of a duality between the non-supersymmetric four
dimensional  $SU(N)$ gauge theory coupled to 6 adjoint scalars fields
and a background of Type $0$ string theory involving a non-vanishing
tachyon field \cite{KT0}.

The presence of such a tachyon instability could seem fatal. However it
 disappears due to a non-perturbative mechanism and the Type $0$ theory
can be used to describe a tachyon-free gauge theory. In this
non-supersymmetric theory the coupling constant is expected to depend
on scale. Nevertheless, the linear logarithmically dependence with the
scale which should appear for the effective coupling constant in
accordance with the short-distance behaviour of the gauge theory
\cite{GW} does not agree with the squared log dependence found for the
effective string coupling \cite{Mi2}, \cite{KT2}. The problem of the
running coupling constant in this theory was also discussed in
\cite{AG} and \cite{Mi3} \footnote{Unfortunately
  we became aware of the  paper \cite{Mi3} only after our paper was
submitted. In the paper \cite{Mi3} besides other things it is shown
that there is a  magnetic screening.}

In order to shed some light to this puzzle we will study in this letter
the leading order for the effective electric and magnetic coupling
constants. In section 2 we describe  a near-horizon geometry in  Type
$0B$ theory
caused by a set of $N$ electric D3-branes. In section 3 we will obtain
 following calculations of Minachan  \cite{Mi2} (see also \cite{KT2})
  the running electric coupling constant $g_e (L)$
 from the  computation of  Wilson loop describing   quark-antiquark  pair. 
  In section 4 we will use a
similar method based on the $DBI$ action for a D-string to study the
 magnetic coupling constant $g_m(L)$ describing monopole-antimonopole
interaction and  obtain the  result that Dirac condition $g_e(L)g_m(L)
= 1$  is scale invariant as it is  expected from
$S$-duality. 

\setcounter{equation}{0}

\section{Near-horizon geometry for the Type $0B$ electrically charged D3-brane}

\ 

    In the near-horizon limit the $IIB$ string theory has the geometry of 
$AdS_5\times
 S^5$ with a self-dual 5-form whose flux through the $S^5$
sphere fixes the charge of the solution. This theory has been proved
to be dual to the 
N=4 supersymmetric $SU(N)$ gauge theory on the boundary $S^4\times S^5$ whose conformal
invariance is due to the  non-running dilaton.

 However, in type $0B$ string theory this conformal invariance is broken. This model has a 
closed string tachyon, all the fermionic partners have been removed and the R-R sector 
doubled. This doubling seems to be crucial for the theory to describe a
tachyon-free gauge theory. This theory also have D-branes
\cite{0branes}. The coupling of the tachyon field $T$ to the
$R-R$ $n-$form gauge field strength shifts the effective mass of $T$
stabilizing in such a way the supergravity background \cite{KT0}.

The ansatz for the metric corresponding to the type $0B$ electrically
charged D3-brane background proposed 
in \cite{KT0} is the following
\begin{equation}
\label{met1}
ds^2= e^{{1\over 2} \Phi} \left( e^{{1\over 2}\xi-5\eta} d\rho^2+ e^{-{1\over 2}\xi} dx_{||}^2
+ e^{{1\over 2}-\eta}d\Omega_5^2\right),
\end{equation}
where $\Phi$, $\xi$ and $\eta$ are functions of $\rho$, parameter related to 
the radial direction $u$ transverse to the 3-brane world-volume $x_{||}$.
 Then the action for the model can be described by a Toda mechanical system
 \begin{equation}
 \label{Toda}
 S=\int d\rho \left[ {1\over 2}\dot{\Phi}^2+{1\over 2}\dot{\xi}^2-
 5\dot{\eta}^2-V(\phi,\xi,\eta)\right]
 \end{equation}
 
 \begin{equation}
 \label{V}
 V= M^2 e^{{1\over 2}\Phi+{1\over 2}\xi-5\eta} + 20 e^{-4\eta}-Q^2f^{-1}(T)e^{-2\xi}.
 \end{equation}
 In this potential the first term comes from the tachyon mass term $M^2={1\over 2} T_{vac}^2$,
 the second one represents the curvature of $S^5$ and the third one is due to the electric $R-R$ charge. There 
  exists also a zero-energy constraint
  \begin{equation}
  \label{tc}
  {1\over 2}\dot{\Phi}^2+{1\over 2}\dot{\xi}^2-5\dot{\eta}^2+V(\phi,\xi,\eta)=0
  \end{equation}
with $Q$ the  charge of the brane system proportional to the number of branes $N$, $T$ the tachyon field and
$f(T)$ a function given by
\begin{equation}
\label{ft}
f(T)= 1+T+{1\over 2} T^2 + O(T^3).
\end{equation}
For a large number of branes, the mass of the tachyon can be neglected and the tachyon can be shown to be a
maximum of the function (\ref{ft}), $f(T)^\prime=0$. Then,
\begin{equation}
\label{ta}
T=T_{vac}= -1, \ \ \ f(T)={1\over 2}.
\end{equation}
It is clear that if $T=0$ we get the constant dilaton and decoupled $\xi$ and $\eta$ fields
\begin{equation}
\label{to}
\Phi=\Phi_0, \ \ e^\xi=2Q \rho, \ \ e^\eta=2\rho^{1/2}, \ \ \rho={1\over u^4}
\end{equation}
which leads to the standard $R-R$ D3-brane solution \cite{3HS}. However,   when $T$ is 
nonzero we can just find approximate solutions.  In the $UV$ the dilaton is expected 
to be slowly varying, at least compared to $\xi$ and $\eta$. Under
this assumption, Minahan \cite{Mi2} solved the equations of motion of
(\ref{Toda}) for $\xi$ and $\eta$ and found
\begin{equation}
\label{xieta}
e^\xi=C_1 \rho, \ \ e^\eta=C_2\rho^{1/2},
\end{equation}
 $C_1=2 Q$ and $C_2=2$ being two constants in a first order
approximation. Then, using (\ref{xieta}) as inputs he obtained that 
\begin{equation}
\label{cc1}
\exp\left({\Phi\over 2}\right)={C_0\over \log{\rho/\rho_0}},
\end{equation}
with $C_0=-8 C_2^5/(T^2\sqrt{C_1})$, is a leading order solution of the
equation of motion for the dilaton. The integration constant $\rho_0$ is
assumed to be $\rho_0>>1$ to assure that the gauge theory length scale
is much greater that the string scale. From  (\ref{cc1}) we can see that the
gauge theory coupling constant $1/g_{YM}$ is no longer constant and
depends on the energy. Thus, if we take the gauge coupling constant as 
$1/g= 4\pi /g_{YM}^2$, then its leading order behaviour is
\begin{equation}
\label{cc2}
{1\over g_{YM}^2}= e^{-\Phi}= 2^{-12}Q(\log{u/\epsilon})^2
\end{equation}
where it has been set $\rho=u^{-4}$ and $\epsilon << u$ is a lower limit
for the energy. Then, the coupling constant runs but with the unexpected
power two for $\log{u/\epsilon}$ instead of the linear dependence which
appears for the effective coupling constant in QCD.

 Finally the solution (\ref{cc1}) can be
used to computed the leading order correction to $C_1$ and $C_2$
\begin{equation}
\label{c1c2}
C_1=2Q \left(1+{1\over \log{u/\epsilon}}\right) \ \ C_2=2 \left(1+{1\over
\log{u/\epsilon}}\right)
\end{equation}
 and the metric in the large $u$ limit {\footnote{In our units
$\alpha^\prime=1$}}
\bea
ds^2= {2^6\over \sqrt{2Q} \log{u/\epsilon}} \left[
(1-{1\over 8 \log{u/\epsilon}}) \frac{u^2}{Q} dx_{||}^2+  \right. \nonumber \\
\left. (1-{9\over 8 \log{u/\epsilon}})\frac{du^2}{u^2} + 
 (1-{1\over 8
\log{u/\epsilon}})d\Omega_5^2 \right].
\label{met2}
\eea
In the Einstein frame, $ ds_s^2=e^{\Phi/2}ds_E^2$, the metric (\ref{met2}) corresponds  to
$AdS_5\times S^5$ space in the large $u$ limit but the lower the energy the more important 
the corrections are and the curvature of $S^5$ becomes smaller than that of $AdS_5$ 
leading to a 10 dimensional space with negative total curvature \cite{Mi2},\cite{KT2} 
{\footnote{ Higher order corrections for the Einstein metric can be found in \cite{KT2}.}}.

As we are interested in the high energy limit we will consider in the
following that the metric in the Einstein frame is still that of the $AdS_5\times S^5$
space but in the string frame we have a running dilaton
\begin{equation}
\label{met3}
ds^2= {2^5\over Q \log{u/\epsilon}} \left[  {u^2\over 2 R_0^2}
  dx_{||}^2+ R_0^2 u^2
du^2 +  R_0^2 d\Omega_5^2  \right],
\end{equation}
$R_0^2=\sqrt{2Q}$ being the radius of $AdS_5$. Then, we are neglecting
any sub-leading correction which scales as $(\log{u/\epsilon})^{-n}$.

\setcounter{equation}{0}

\section{The quark-antiquark interaction}

\

A pair of massive quark-antiquark in the background of $N$ electric D3-branes
 can be realized as a string  starting and ending on a D3-brane which has 
been separated an infinity distance from the set of $N$ branes \cite{M2}\cite{RY}.
This separation breaks the group $U(N+1)$ to $U(N)\times U(1)$ by giving an expectation 
value $<H>$ to a Higgs field. The massive W-bosons which appear have a
mass proportional to $<H>$ and transform in the fundamental
representation of $U(N)$. They will play the role of quarks. 

Denote the string coordinates by $X^n(\tau,\sigma)$ where $\tau$, $\sigma$ parameterize
 the string  world-sheet, then the action for the string is
\begin{equation}
\label{stac}
S= {1\over 2 \pi} \int d\tau d\sigma \ \sqrt{ \det G_{mn} \partial_\alpha
X^m\partial_\beta X^n},
\end{equation}
 $G_{mn}$ being the Euclidean metric in (\ref{met3}). Since we are
interested in a static configuration independent of the $\Omega_5$
modes, we take $\tau=t$, $\sigma=x$ where $x$ is a direction along the
three-branes. The action simplifies to
\begin{equation}
\label{qact}
S={1\over 2\pi}\int dx \ e^{\Phi/2}\sqrt{{U^4\over 4 R_0^2} X^{\prime
2}+ {1\over 2}(\partial_x U)^2}
\end{equation}
where $X^\prime=1$, $T$ is a constant due to the integration over $t$ and 
\begin{equation}
e^{\Phi/2}= {2^6\over \sqrt{Q} \log{u/\epsilon}}
\end{equation}

 is the running dilaton we found in 
the previous section. Notice that, although the dilaton did not appear explicitly
 in the original action for the string (\ref{stac}) the running radius
of the $AdS$ space makes it to enter the game. This fact will be relevant in
order to provide the linear log dependence for the gauge coupling
constant.

The action (\ref{qact}) does not show any explicit
dependence on $X$ and that allows us to solve  its Euler-Lagrange
equation of motion
\begin{equation}
\label{qem1}
      \partial_x \left[ e^{\Phi/2} {{U^4\over 4 R_0^4}\over \sqrt{{U^4\over 4
R_0^4}+ {1\over 2}(\partial_x U)^2}}\right] =0.
\end{equation}
Since  we are working in
the approximation of a large value for $\log{u/\epsilon}$, the
 $\partial_x\log{u/\epsilon}$ is a sub-leading
correction and the factor $e^{\Phi/2}$ can be thought of as a constant.
Thus, 
we can approach (\ref{qem1}) to
\begin{equation}
\label{qem2}
     e^{\Phi/2}  \partial_x \left[ {{U^4\over 4 R_0^4}\over \sqrt{{U^4\over 4
R_0^4}+ {1\over 2}(\partial_x U)^2}}\right] =0
\end{equation}
which leads to
\begin{equation}
\label{qem3}
      \left[ {{U^4\over 4 R_0^4}\over \sqrt{{U^4\over 4
R_0^4}+ {1\over 2}(\partial_x U)^2}}\right] = {U_0^2\over 2 R_0^2}
\end{equation}
where the leading order of  $U_0$, the minimum value of $U$, is a constant 
to be determined. If we consider 
the quark placed at  $x=L/2$ and the antiquark at $x=-L/2$ the minimum
value for $U$ occurs at $x=0$ by symmetry. Therefore we can write $x$ as a function of $U$
\begin{equation}
\label{xu}
x=\sqrt{2}{ R_0^2\over U_0}\int_1^y {dy\over y^2\sqrt{y^4-1}} \ \ , \
y=U/U_0 
\end{equation}
where $U_0$ must satisfy the condition
\begin{equation}
\label{uo}
{L\over 2}=\sqrt{2}{ R_0^2\over U_0}\int_1^\infty {dy\over
y^2\sqrt{y^4-1}}= { R_0^2\over U_0}{2\pi^{3/2}\over \Gamma(1/4)^2}.
\end{equation}
Now taking into account (\ref{qem3}), (\ref{xu}) and the value for $U_0$
(\ref{uo}) we can return to (\ref{qact}) to compute the total energy of the
quark-antiquark configuration
\bea
E_{q\bar q}=  {\sqrt{2}U_0\over 2\pi}\int_1^\infty dy \ e^{\Phi/2} { y^2\over 
 \sqrt{y^4-1}} \nonumber \\
 {2^{11/2}U_0\over Q\pi}\int_1^\infty dy \ {1\over \log{(yU_0/\epsilon)}} { y^2\over 
 \sqrt{y^4-1}}.
\label{qen}
\eea

Since $\epsilon$ is an UV cutoff for the energy $\epsilon<<U_0$ we can approach 
\begin{equation}
\label{log}
\log{(yU_0/\epsilon)}=\log{y}+\log{U_0/\epsilon}\sim log{ L_0/L}
\end{equation}
where we have used the relation (\ref{uo}) with $L_0=
{R_0^2\pi^{3/2}\over \epsilon \Gamma(1/4)^2}$  the UV cutoff for the distance $L<<L_0$.
Therefore we have found that the leading contribution to the energy is
\begin{equation}
\label{qen2}
E_{q\bar q}= {2^7\sqrt{Q}\pi^{1/2}\over \Gamma(1/4)^2}{1\over L \log{(L_0/L)}}\int_1^\infty dy \  { y^2\over 
 \sqrt{y^4-1}}.
\end{equation}
This result is infinity because we are including the masses of the 
quarks which correspond to strings stretched from the D3-brane at $U_{max}$ to the 
$N$ D3-branes at $U=0$. Then, by integrating the energy up to  $U_{max}$ we will 
subtract the regularized mass and find  that the remaining energy is finite
\cite{M2} 
\begin{equation}
\label{qen3}
E_{q\bar q}= {2^6 \sqrt{2Q}\pi^2\over \Gamma(1/4)^4}{1\over L \log{(L_0/L)}}
\end{equation}
and, as claimed in \cite{Mi2}, the effective coupling between a heavy quark 
and its antiquark does show the expected linear dependence 
\begin{equation}
\label{ccq}
 {1\over g_{e}^2}\propto \log{(L_0/L)}.
\end{equation}
This result leads to reconsider the relation $1/g= 4\pi/g_{YM}^2$ between the 
string and the gauge coupling constants.

\setcounter{equation}{0}

\section{The monopole-antimonopole interaction}

\

The pair of massive monopole-antimonopole in the background of $N$ electric D3-branes
 can be simulated in a similar way as the quark-antiquark pair of the previous section.
For the monopole-antimonopole configuration we will take a type $0$ electric  D-string 
 with both ends on a D3-brane  separated an infinity distance from the set of $N$ branes.
 The same approach was used in \cite{Mi1} in case of $N=4$
superconformal Yang-Mills theory.
 The world sheet action for the D-string spread in the $01$ plane is the
 Born-Infeld action {\footnote{ A 
tachyon-dependent function $k(T)$ might appear as a multiplicative factor 
\cite{KT0}. However, in case of a constant tachyon it is not likely to play a 
relevant role in the determination of the log dependence of the gauge coupling 
constant.}}
\begin{equation}
\label{dstac}
S= {1\over 2\pi}\int dt dx \ e^{-\Phi}\sqrt{ \det G_{mn} \partial_\alpha
X^m\partial_\beta X^n},
\end{equation}
 $G_{mn}$ being the Euclidean metric in (\ref{met3}). We will suppose
 a static configuration independent of the $\Omega_5$
modes and, in such a way,   this action can be written as follows
\begin{equation}
\label{mact}
S={T\over 2\pi} \int dx \ e^{-\Phi/2}\sqrt{{U^4\over 4 R_0^2} X^{\prime
2}+ {1\over 2}(\partial_x U)^2}
\end{equation}
where $X^\prime=1$ and 
\bea
e^{-\Phi/2}= \frac{\sqrt{Q}\log{u/\epsilon}}{ 2^6}
\eea
 It is remarkable to 
notice that this action differs from that of the quark-antiquark (\ref{qact}) just in the 
factor $ e^{-\Phi}$. This factor will be crucial to obtain a different $\log$ dependence 
for the monopole effective gauge coupling constant. Again the running
radius of the $AdS$ space changes the dilatonic factor of the action
providing, as we will see, the linear log dependence for the effective magnetic 
coupling.

The action (\ref{mact}) does not depend explicitly on $X$ either and the Euler-Lagrange
 equation of motion for this variable reads
\begin{equation}
\label{mem1}
      \partial_x \left[ e^{-\Phi/2} {{U^4\over 4 R_0^4}\over \sqrt{{U^4\over 4
R_0^4}+ {1\over 2}(\partial_x U)^2}}\right] =0.
\end{equation}
We can suppose again that the factor $e^{-\Phi/2}$ has a slow variation and use 
\begin{equation}
\label{mem3}
      \left[ {{U^4\over 4 R_0^4}\over \sqrt{{U^4\over 4
R_0^4}+ {1\over 2}(\partial_x U)^2}}\right] = {U_0^2\over 2 R_0^2}
\end{equation}
to be able to write $x$ as a function of $U$ (\ref{xu}) and to fix the value of 
 $U_0$ (\ref{uo}). Hence the minimum of the energy for the monopole-antimonopole
 configuration happens to be the same constant as that of the quark-antiquark 
case in a first approximation. However, from (\ref{qem1}) and (\ref{mem1}) we 
expect them to differ due to higher order corrections.
 
Now we can plug the results (\ref{xu}) and (\ref{uo}) into the expression
(\ref{mact}) to obtain the monopole potential
\bea
E_{q\bar q}=  {\sqrt{2}U_0\over 2\pi}\int_1^\infty dy \ e^{-\Phi/2} { y^2\over 
 \sqrt{y^4-1}}= \nonumber \\ 
  {2^{13/2}U_0\over Q}\int_1^\infty dy \  \log{(yU_0/\epsilon)} { y^2\over 
 \sqrt{y^4-1}}.
\label{men}
\eea
Here, as in the previous section, taking into account that $\epsilon$ is an UV 
cutoff for the energy $\epsilon<<U_0$ we  can use (\ref{log}) to get
the leading infinite contribution to the energy 
\begin{equation}
\label{men2}
E_{m\bar m}= {2^7\sqrt{Q}\pi^{1/2}\over \Gamma(1/4)^2}{\log{(L_0/L)}\over L }\int_1^\infty dy \  { y^2\over 
 \sqrt{y^4-1}}.
\end{equation}
In this case, the infinity is due to the masses of the 
monopoles, D-strings joining the D3-brane at $U_{max}$ to the 
$N$ D3-branes at $U=0$. The integration the energy up to  $U_{max}$  will
 allow us to  subtract the regularized mass  and obtain a finite result
\begin{equation}
\label{men3}
E_{m\bar m}= {2^6 \sqrt{2Q}\pi^2\over \Gamma(1/4)^4}{\log{(L_0/L)}\over L }.
\end{equation}
Therefore, we have found that the effective coupling between a heavy monopole 
and antimonopole  is of zero-charge type and shows the inverse $\log$ dependence 
\begin{equation}
\label{ccm}
 {1\over g_{m}^2}\propto {1\over\log{(L_0/L)}}
\end{equation}
compared to ${1\over g_{e}^2}$ (\ref{ccq}). Hence the electric and magnetic 
coupling constants verify the Dirac condition
\begin{equation}
\label{dirac}
g_e (L)g_m (L) =1
\end{equation}
which has not been destroyed by  running coupling constant.

\section{Conclusion and Acknowledgments }

\

The  dual gravity description of
non-supersymmetric gauge theories  has been used 
in this paper to  check that Dirac relation between electric and
magnetic charges  is RG invariant (at least
in a logarithmic approximation). It seems that this
 is a necessary element for the
self-consistency of the theory. It was also shown that in this theory the
running coupling constant is rather  proportional to $ \exp(-\Phi/2)$
than to $\exp(-\Phi)$ as one could expect. The fact that there is a
square of logarithm in  $\exp(-\Phi)$ was crucial to obtain the
correct renormalization of the magnetic coupling constant. We did not
answer the question why the coupling constant is square root of what
we might expect naively and this seems to be open important question.
 It will be nice to find running coupling constant in this theory
using other approaches, for example studying instanton effects and we
plan to return to this issue in  the future.

We are grateful to A. Tseytlin and  M. Gaberdiel for useful
conversations.  The work
of G.L. was supported by the spanish FPU programme under grant
FP9717442117.

\end{document}